\documentstyle[12pt,epsf]{article}

\begin{document}
\begin{sloppypar}
\begin{center}

{\bf GLUON CORRELATION MOMENTS RATIO IN THE INSTANTON FIELD \\ }
\vspace{15pt}
{\small V.KUVSHINOV AND R.SHULYAKOVSKY      \\

{\it Institute of Physics, National Academy of Sciences of Belarus       \\
Minsk 220072 Scaryna av.,70 \\
E-mail: kuvshino@dragon.bas-net.by and \\
shul@dragon.bas-net.by }} \\

\vspace{15pt}
\end{center}
\noindent
The instanton-induced multiple events in high energy collisions
are considered in nonperturbative quantum
chromodynamics (QCD).
Here we obtained unusual behaviour of ratio of correlation moments
$H_q$ for such processes which can be used for
experimental search of instantons.

\vspace{15 pt}
\noindent
As it is known, Yang-Mills gauge theories
have highly degenerated vacuum structure on the classical level~[1].
Quantum tunnelling transitions between
different vacuum states are associated with instantons~[2].

Experimental search of the QCD-instantons goes already
at HERA (DESY, Hamburg) in electron-proton
deep inelastic scattering~[3].
There are following theoretically predicted features of
instanton canal of multiple production:
high parton multiplicity (about $10\div 20$ at HERA~[3,4]);
isotopically parton distribution in the instanton rest system and
homogeneous quarks flavours distribution~[5];
specific behaviour of total cross section~[4] and
two-particle correlation function~[6].

Here we study the behaviour of ratio of correlation moments
$H_q=~K_q/F_q$~[7]
as the new criterions of instanton identification.
Here $F_q$, $K_q$ and $H_q$
are factorial, cumulant and co-called $H_q$-moments correspondingly.
Correlation moments ratio $H_q$ is {\it more
precise} quantity for distinguishing of multiplicity distribution~[7].

In quasiclassical approximation Poisson distribution for the
probability of $n$ gluon production was obtained
for the instanton-induced multigluon final states~[3,8].
In this case we have the trivial results:
$G(z)=e^{A[z-1]}$, $F_q=1$, $K_q=\delta_{q1}$, $H_1=\delta_{q1}$.

Taking into account first quantum correction we obtain
the following formula for the generating function takes place~[6]:

$$
G(z)\equiv\sum\limits_{n=0}^{\infty}P_nz^n
=e^{A[z-1]}\frac{1+Bz^2}{1+B},\quad
A=\frac{4\pi}{\alpha_0}\biggl(\frac{1-x'}{x'}\biggr)^2,
$$
$$
B=-\frac{2\pi }{\alpha_0}\biggl(\frac{1-x'}{x'}
\biggr)^3,\quad x'\sim 0.5\div 1.
\eqno(1)
$$

\noindent
where coupling constant of strong interaction
$\alpha_0=\alpha (\rho_{cut})$, $\rho_{cut}$ is instanton size cut off,
$x'$ -- Bjorken variable of parton-parton collisions.

By straight calculation we obtain $H_q$ moments behaviour
as a function on $q$ (Fig. 1).
$H_q$-moments are negative, have first minimum at
$q=2$.
Such dependence of $H_q$ on their rank $q$ may be new criterion of
identification of instantons at experiment.

\hspace{2cm}
\begin{minipage}{6cm}
\epsfxsize=3.45in \epsfbox{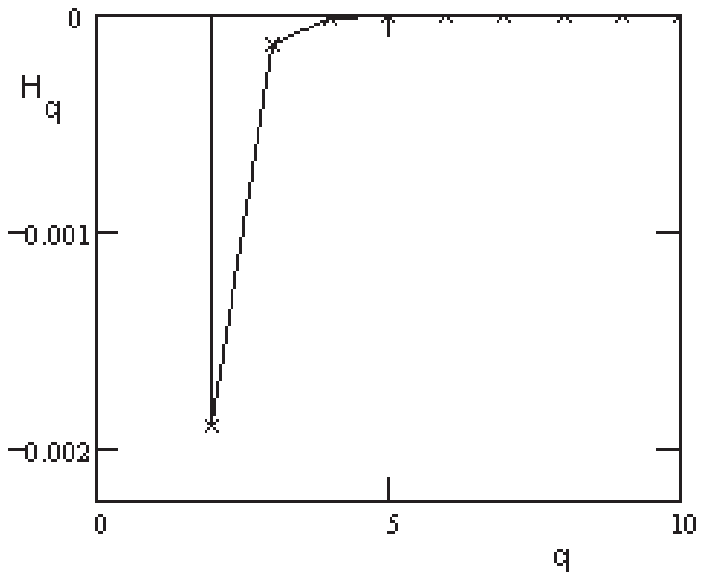}
\end{minipage}

\noindent
\vspace{-1.5cm}
\begin{center}
Fig. 1. $H_q$-moments as the functions of their ranks.
\end{center}

\vspace{12pt}
The reason is unusual and specific position of the first minimum
for this nonperturbative process, which doesn't move
by the next quantum corrections in chosen interval of
variables as estimations show.
Perturbative QCD calculations confirmed by experimental data
for ordinary multiple production of different types give the first
minimum at $q=5$.
Such clear distinction of perturbative and nonperturbative
calculations is of principle both from experimental and theatrical
point of view.

The authors are grateful for the support in part to
Basic Science Foundation of Belarus
(Projects M96-023, F97-013).

\end{sloppypar}
\end{document}